# Verifying the Medical Specialty from User Profile of Online Community for Health-Related Advices

Solomiia Fedushko[0000-0001-7548-5856], Natalia Shakhovska[0000-0002-6875-8534], and Yuriy Syerov[0000-0002-5293-4791]

Lviv Polytechnic National University, Lviv, Ukraine

`solomiia.s.fedushko@lpnu.ua, nataliya.b.shakhovska@lpnu.ua,`
`yurii.o.sierov@lpnu.ua`

**Abstract.** The paper describes the verifying methods of medical specialty from user profile of online community for health-related advices. To avoid critical situations with the proliferation of unverified and inaccurate information in medical online community, it is necessary to develop a comprehensive software solution for verifying the user medical specialty of online community for health-related advices. The algorithm for forming the information profile of a medical online community user is designed. The scheme systems of formation of indicators of user specialization in the profession based on a training sample is presented. The method of forming the user information profile of online community for health-related advices by computer-linguistic analysis of the information content is suggested. The system of indicators based on a training sample of users in medical online communities is formed. The matrix of medical specialties indicators and method of determining weight coefficients these indicators is investigated. The proposed method of verifying the medical specialty from user profile is tested in online medical community.

## 1    Introduction

Available research methods are reduced to a fragmentary solution to the problem, they are theoretical and the results of these studies are mostly not tested in practice.
Increasingly, people are turning to online communities for health-related advices. This process has both positive and negative factors. Positive: anonymity, response time, variety of thoughts and information. Negative: falsehood, incompetence, commercial interest. Published information in the online community for health-related advices can both help and harm.

The community administration is responsible for the advice that patients receive. And it is important for community administrator to verify medical specialty from user profile and identify persons that state that they are specialists but give incompetent and inappropriate advice. Because of given reasons, developing a method of verifying the medical specialty from the user profile of the online community for health-related advices is an actual and important task.

Available research methods are reduced to a fragmentary solution to the problem, they are theoretical and the results of these studies are mostly not tested in practice.

## 2      Analysis of related research

An analysis of the activities of online communities is the subject of research, among which are clearly distinguished in web content mining. One of the important problems of analyzing the content of online communities is the analysis of the user personal data. Despite the significant importance for further development of this research area, no effective methods for analyzing *personal information in the profile of the online community user have yet been developed* (see in Fig.1).

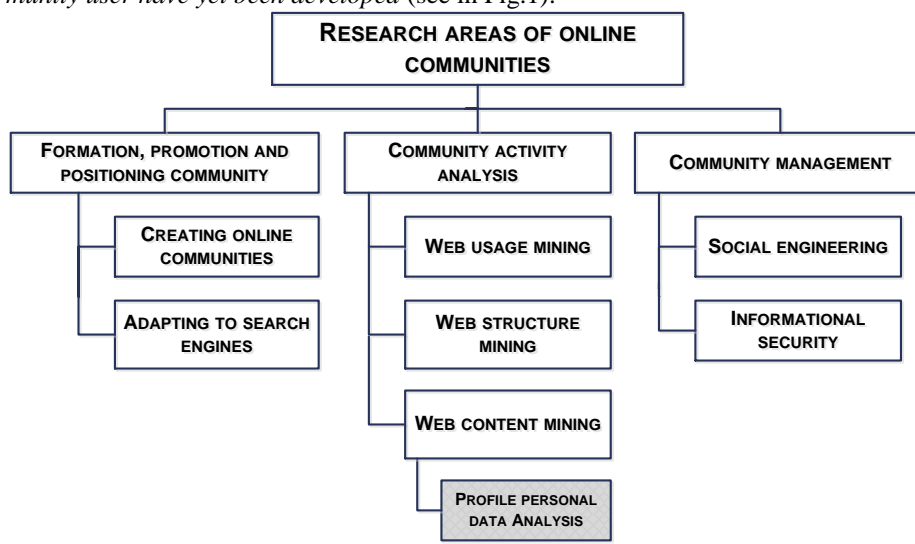

**Fig. 1.** Scheme of research areas of online communities

Verification of the probability of personal information placed in the WWW system is an actual and important subject of scientific research in such areas:

- certainty and quality of medical personal data [1, 2];
- validation of online information, strategies for assessing the reliability and usefulness of medical information on the Internet [3-5];
- definition and justification of validity and reliability (in particular, content), relevance of the content [6];
- accuracy and reliability of internet sources of specific (educational, military, scientific) web content [7-10], highly specialized content [11];
- verification of the profile personal data of specialized online community (in particular the medical online communities [12-15]).

The latest research direction is the least developed. In this area, research is dedicated to verifying the personal data of the medical online community users.

The results of scientific research in this area are in demand by a wide range of specialists in the organization and functioning of medical online communities, as those that should ensure their success and effectiveness.

Considering of the above analysis of scientific works, among the well-known literary sources, there is a lack of thorough research on the verification of personal information of user medical profiles and the study of the reliability of personal data of users in social communications, in particular, medical online communities, in order to improve their functioning. This, in turn, generates the actual problem of developing new methods and tools to analyze the reliability of the user personal data which would have adequate scientific justification, formalization, predictable efficiency and versatility.

## 3     Methods of research

### 3.1    Formation of a system of indicators based on a training sample of users in medical online communities

The functioning of the system of formation of indicators consists in the creation and processing of informational content of a training sample of online community users. The scheme of the formation of indicators based on a training sample of users in online community is given on Fig.2.

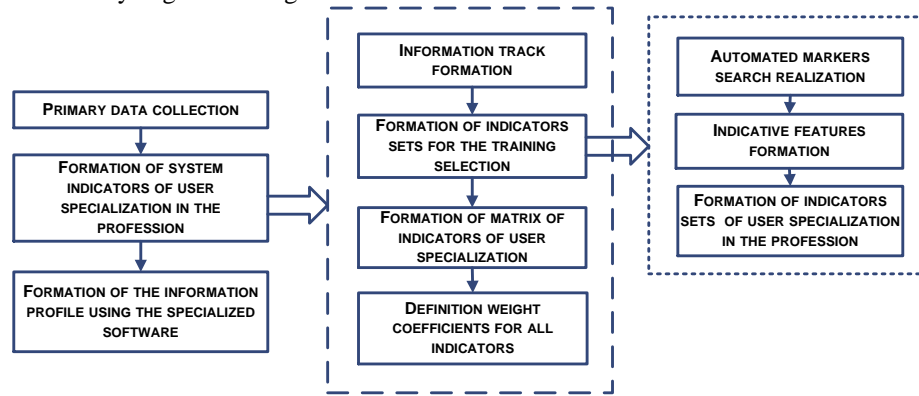

**Fig. 2.** Scheme systems of formation of indicators of user specialization in the profession based on a training sample

Now let's consider each of the stages of the functioning of the system of formation of indicators based on a training sample of medical online community:

- Stage I. Primary data collection
- Stage II. Formation of the system of indicators
- Stage III. Formation of an information profile using the using specialized software

### 1.2.1. Primary data collection.

Primary data collection is only performed from trusted sources and trusted users of medical online communities. The reliability of information sources for research is a decisive factor in obtaining a true and credible result. To this end, the collection of primary data from the online community administrators has been carried out. The information content is with a high degree of truthfulness, is selected to create the training sample. The administrator or moderator is personally familiar with users of the online community verified by time.

### 1.2.2. Formation of the system of indicators

At this stage, sets of linguistic and communicative indicators are formed by automated analysis of the information track of user of the medical online community.
The formation of the information track is carried out in accordance with the model of the information track of the member of the online community.

*Formation of indicator sets for the training sample.* According to the structural model of the indicators of the user specialization of the medical online community, the division of the online community user according to each studied.

Specialization of profession of the group is chosen. Each web-user for this training sample is carefully selected, considering the reliability of the user personal information in the medical online community and the reliability of the information track of the user of the medical web forum.

It is also taken into account the fact that the results of research significantly affect both the context of the messages and the topics of discussions. Taking into account this fact, the basis of this research is a diverse selection of information track of users from all thematic sections, two Ukrainian-language web forums.

The definition of the features of Internet communication was carried out by analyzing information track of more than three thousands users of Ukrainian-speaking medical online community. Analysis of the informational track of user of Ukrainian-language web forums for the presence of grammatical, lexical-semantic and lexical-syntactic features that is more closely related to the medical specialization of the online community users is conducted.

The formation of sets of indicators for a training sample consists in the following steps:

1. Automated search for markers
2. Formation of indicative features.
3. Formation of sets of indicators.

The main task of this process is consolidation of indicative features of online communication. The formation of sets of indicators consists in grouping indicative features into intuitive semantic groups.

To find and correct errors of the set scientists developed many algorithms for English and Ukrainian, although not as thoroughly as it is needed. In this regard, the development of a new automated tool for finding and correlating errors in web content is not decisive. Since the best solution to this problem is an existing well-functioning automated tool, which is the analysis of text filtering words, the selection of words with errors and their correction.

**Formation of the matrix of indicators.**
Based on sets of indicators, experts form the matrix of indicators (1) by the method of computer-linguistic analysis of the information content of online communities for each value of the medical specialization of a particular user, which we define separately. As a result, for each value of a certain medical specialization is obtained a matrix of indicators:

$$Indicator^{(MedSp,OC)} = \begin{pmatrix} Ind_{1,1}^{(MedSp,OC)} & \cdots & Ind_{1,j}^{(MedSp,OC)} & \cdots & Ind_{1,N\_Vl(MedSp,OC)}^{(MedSp,OC)} \\ \cdots & \cdots & \cdots & \cdots & \cdots \\ Ind_{i,1}^{(MedSp,OC)} & \cdots & Ind_{i,j}^{(MedSp,OC)} & \cdots & Ind_{i,N\_Vl(MedSp,OC)}^{(MedSp,OC)} \\ \cdots & \cdots & \cdots & \cdots & \cdots \\ Ind_{N\_Ind(MedSp,OC),1}^{(MedSp,OC)} & \cdots & Ind_{N\_Ind(MedSp,OC),j}^{(MedSp,OC)} & \cdots & Ind_{N\_Ind(MedSp),N\_Vl(MedSp,OC)}^{(MedSp,OC)} \end{pmatrix} \quad (1)$$

where $N\_Vl$ is a function that for each medical specialization determines the number of values of this personal data; $N\_Ind$ is a function that for each value of the medical specialization determines the number of indicators of this value of medical specialization. Each line of the matrix (2) is a vector of indicators of a certain medical specialization:

$$Ind^{(MedSp,OC)} = \left( Ind_{1,1}^{(MedSp,OC)} \quad \cdots \quad Ind_{N\_Ind(MedSp,OC),j}^{(MedSp,OC)} \quad \cdots \quad Ind_{N\_Ind(MedSp),N\_Vl(MedSp,OC)}^{(MedSp,OC)} \right) \quad (2)$$

The matrix column (3) is a vector of indicators for a certain value of the medical specialization of the investigated online community:

$$Indicator^{(Med,OC)} = \begin{pmatrix} Ind_{1,1}^{(Med,OC)} \\ Ind_{i,1}^{(Med,OC)} \\ Ind_{N\_Ind(Med,OC),1}^{(Med,OC)} \end{pmatrix} \quad (3)$$

Based on this principle, we create a matrix for each web user.

To calculate the distance from the reference value of the medical specialization to each possible value of the medical specialization of the atomic k-th user of the online community, we use the formula for determining the Euclidean distance as the basis:

$$\rho_j^{(k)}(Value, User) = \sqrt{\sum_{i=1}^{N\_Ind(MedSp,OC)} \left( Ind_{i,j}^{(MedSp,OC)} - Ind_{i,j}^{(MedSp,U)} \right)^2 * w_i^{(MedSp)}} \quad (4)$$

when $k \in 1 \ldots N\_Vl(MedSp,OC)$; $w_i^{(MedSp)}$ – weight coefficient of a specific indicator of a specific value of medical specialization.

As a result, we select the value of the user medical specialization for which it is valid. Moreover, the matrix is universal for all values of a specific medical specialization of a specific online community for which the models are synthesized. Depending on the subject and type of the online community, a model for each of the values of the medical specialization is synthesized using an automated information-analytical monitoring system. Weighted coefficients of indicators are presented in the vector:

$$W^{(Vl,MedSp)} = \left( w_1^{(Vl,MedSp)} \quad \ldots \quad w_j^{(Vl,MedSp)} \quad \ldots \quad w_{N\_Ind(MedSp,Vc)}^{(Vl,MedSp)} \right) \quad (5)$$

Vector of weight coefficients of indicators of user medical specialization of the value of *MedSp-Vl* is obtained as a result of the work of automated information-analytical monitoring system.

*The importance of indicators is determined by the weighting coefficients.*

The results of the analysis vary according to the specifics of the online community. The larger the coefficient, the more important is the lingua-communicative indicator for verifying the corresponding user medical specialization in a particular online community.

*Determination of weight coefficients indicators.*

The determination of weight coefficients for indicators of all values of user medical specialization takes place using the information system of multilevel computer monitoring. At the stage of forming an array of input data of the multi-level monitoring information system, the information track of the users of the online communities are processed for the presence of their markers in order to form sets of indicators for a particular online community with the relevant subject.

### 3.2 Forming the user information profile of online community for health-related advices

The validation of the web personality is a complex process of determining the basic personal data of the online medical community. Validation of online users' data is not only a process of identifying online users, but also an important point in verifying the authenticity of the user personal information of online medical community and categorizing the users in accordance with the level of reliability of their personal data.

This non-trivial task requires the development of special software (see Fig.3) for the information tracks building of online medical community users.

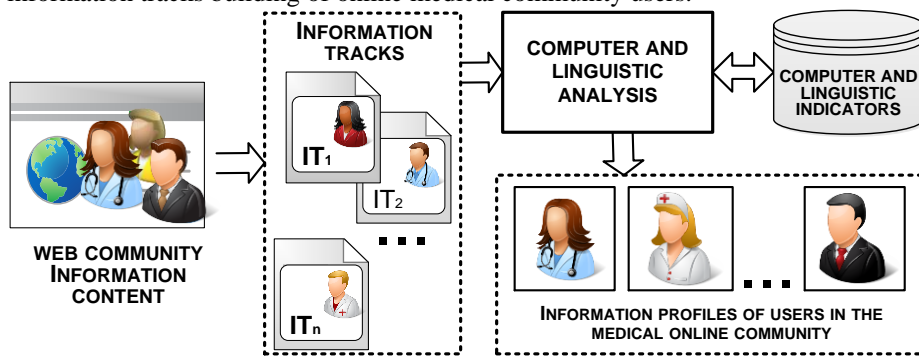

**Fig. 3.** Scheme of the process of formation the user information profiles of online community for health-related advices

The initial result is user information profiles of online community for health-related advices that formed on the basis of computer-linguistic analysis of the information track of the users of the online medical community. This allows verifying the medical specialty from user profile of online community for health-related advices. A software tool for validating medical specialty of online users based on the method of computer-linguistic analysis of information tracks of users of medical online community.

Information profile of users of medical online community is built only from the verified user personal data by the method of computer-linguistic analysis of content of online community for health-related advices. So, process of building the information profile includes also data of medical specialty.

The information profiles are performed in accordance with the developed algorithm for forming the information profile of a medical online community user. The diagram of this algorithm is shown in Fig. 4. The purpose of the algorithm for forming the user information profile of online community for health-related advices is to verify the maximum amount of personal information that a user of the online medical community has specified in account by computer-linguistic analysis.

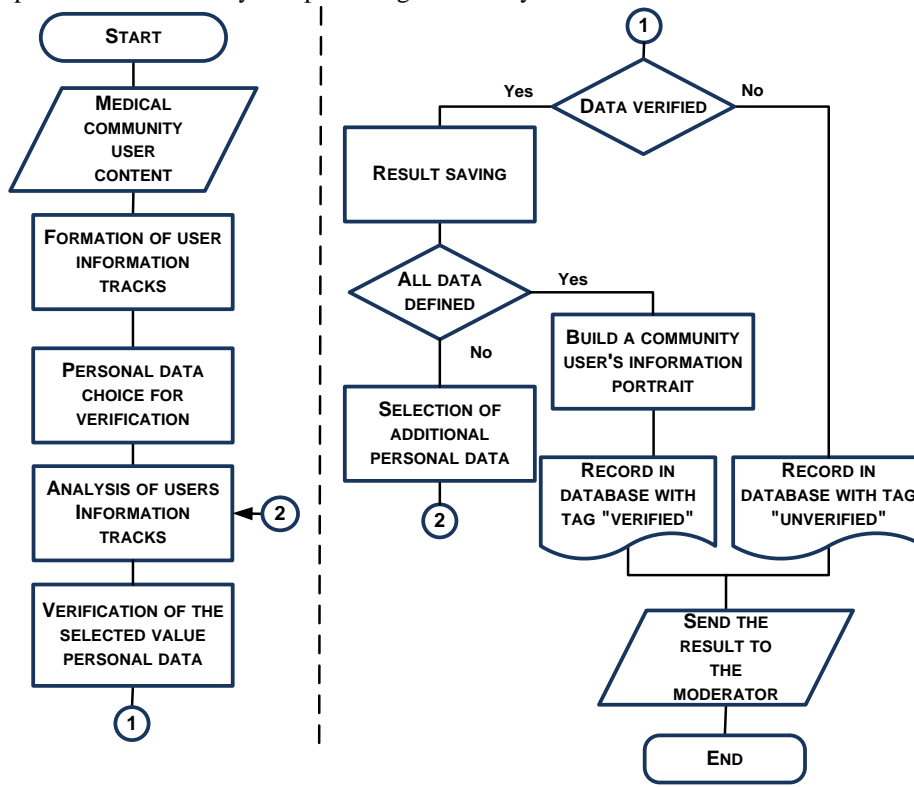

**Fig. 4.** Diagram of the algorithm for forming the information profile of a medical online community user

The basis of the algorithm is the informational track of user information profiles of online community for health-related advices. The formal model of the information track of a user of online community for health-related advices is described in the previous works [16]. The result of developed algorithm is classification of information profile of online medical users.

## 4   Results and Discussion

The developed method of verifying the medical specialty from user profile of online community for health-related advices is used for analysis medical specialties of users of online medical community "Ukrainian doctors forum" [17]. As shown in Fig. 5, the system is processed by an average 1/3 part of users who provide professional advice on the specialty in their profile. These users have created enough content to form their information track and verify their specialty. Users, who is registered, but do not have established the minimum amount of content in the online medical community for verification automatically assigned to a group of users with unverified medical specialty.

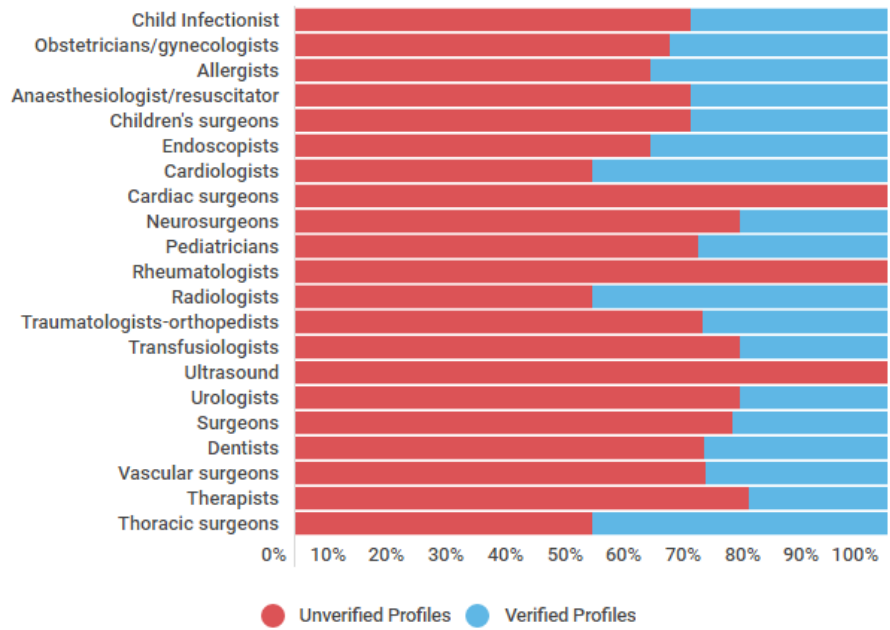

**Fig. 5.** The diagram of verification results of medical specialties of profile users of online medical community

The results of testing proposed verification methods on user profiles of online medical community "Ukrainian doctors forum" is definition 25.75% user profiles with verified medical specialties of users which are filled the field of medical specialties in own profile. Also online medical community "Ukrainian doctors forum" consist of 49.16% users with medical specialties, 5.69% of all community users with non-medical specialties, 25.96% of profile with missing data about medical specialties and 4.65% of all users filled in community user profile incorrect data about specialties.

## 5 Conclusion

The paper presents the verifying methods of medical specialty from user profile of online community for health-related advices. Implementation of the developed methods in medical online communities will contribute to the formation of a new vision and perception of specialized online communities in Ukrainian society.

After all, every user of the Internet from time to time needs a competent response to an urgent question about their health situation or their relatives. Exactly the competence of the answers by experts with a certain specialization is a critical issue. Developed method enables the owner of online medical communities to identify the users who provide competent health-related advices.